# on Dynamical Behaviors in the Fractional Generalized Langevin Equation

Yun Jeong KANG[1] and Kyungsik KIM[2,*]

[1]*College of Liberal Studies and College of Convergence Liberal Arts, Wonkwnag University, 54538, Korea*
[2]*Department of Physics, Pukyong National University, Busan 48513, Korea*

**Abstract**
The focus of our study in this paper is on the active dynamics and a fractional generalized Langevin equation with a memory kernel $K(t) \sim t^{2H-2}$. The Fokker-Planck equation is obtained by deriving it from a second-order differential equation. The joint probability density we obtained enables us to calculate various statistical quantities such as mean squared displacement and mean squared velocity in different three-time domains.

Researchers have reently been extensively studying the nonequilibrium motion of active particles in different models such as run-and-tumble model[1,2], active Brownian particle[3,4], active Langevin particle[5], active Ornstein-Uhlenbeck particle[6-8], self-propelled Janus colloids[9,10], and biological swimmers[11].

The active diffusion, analyzed through theories, computer simulations, and experiments with active viscoelastic systems, showed similarities and differences across the board. Examples includes the transport of passive tracer in active baths and cells[12], chromosomal dynamics[13], lateral diffusion of membrane proteins[14], and the tracer diffusion in dense colloidal systems[15]. The active particles have studied in polymeric environments such as the actomyosin and endoplasmic reticulum networks[16], the microtubule, and the macromolecule bound to polymer strands such as the DNA and the chromosome[17].

The active fractional Langevin equation, a new class of nonequilibrium dynamic models, is presented; its unique viscoelastic memory effect is detailed, characterized by a power-law decay of correlations over time. We will derive a Fokker-Planck equation for the joint probability density from the fractional generalized Langevin equation with the thermal equilibrium noise. A second-order differential equation (fractional generalized Langevin equation) in our model are expressed in terms of

$$\frac{d^2}{dt^2}x(t) = -\gamma \int_0^t dt'(t-t')^{2H-2}v(t') + \eta_H \zeta_H(t). \quad (1)$$

Here, a memory kernel denotes $K(t-t') = (t-t')^{2H-2}$ and the thermal equilibrium noise denotes $\zeta_H(t)$. We add the random force as

$$<\zeta_H(t)\zeta_H(t')> = \eta_H^2 \zeta_H(t-t') = \frac{\eta_H^2}{2\tau_H}\left|\frac{t-t'}{\tau_H}\right|^{2H-2}. \quad (2)$$

Here, the thermal energy is $k_B T$, the correlation time $\tau_H$, and $\eta_H^2 = 2\gamma k_B T$.

The joint probability density $p(x,v,t)$, for the displacement $x$ and the velocity $v$, is defined by

$$p(x,v,t) = <\delta(x-x(t))\delta(v-v(t))>. \quad (3)$$

By taking time derivatives in the joint probability density and inserting Eq. (1) into Eq. (3), we can write the time derivative of the joint probability equation for $p(x,v,t) \equiv p$, $\delta_x(x-x(t)) \equiv \delta_x$, and $\delta_v(v-v(t)) \equiv \delta_v$ in the form:

$$\frac{\partial}{\partial t}p = -\frac{\partial}{\partial x}<\frac{\partial x}{\partial t}\delta(x-x(t))\delta(v-v(t))>$$
$$-\frac{\partial}{\partial v}<[-\gamma\int_0^t dt'(t-t')^{2H-2}v(t') + \eta_H\zeta_H(t)]\delta_x\delta_v>. \quad (4)$$

Here, the fractal dimension has $1/2 < H < 1$. In the above expression $-\gamma \int_0^t dt'(t-t')^{2H-2}v(t') = -\gamma[\Gamma(2H-1)]^{-1}\frac{d^{2-2H}}{dt^{2-2H}}x(t)$, and we put $\Gamma(2H-1) = 1$. We assume from the joint probability density that the particle is initially at rest at time $t = 0$. Then the joint probability densities are derived as

$$\frac{\partial}{\partial t}p = [-v\frac{\partial}{\partial x} + \gamma D_{2-2H}\frac{\partial}{\partial v}x]p - a[\frac{t^{2H}}{2H\tau_H^{2H}}\frac{\partial^2}{\partial v\partial x} + \frac{t^{2H-1}}{(2H-1)\tau_H^{2H-1}}\frac{\partial^2}{\partial v^2}]p, \quad (5)$$

where $D_{2-2H} = d^{2-2H}/dt^{2-2H}$ and $a = \eta_H^2/2$. It is apparent that Eq. (5) is called the Fokker-Planck equation. In Ref. [18], some derivations in relation to the correlated Gaussian force are given and used. As we define the double Fourier transform of the joint probability density $p(\xi,v,t)$ by the equation

$$p(\xi,v,t) = \int_{-\infty}^{+\infty}dx\int_{-\infty}^{+\infty}dv\exp(-i\xi x - ivv)p(x,v,t), \quad (6)$$

the Fourier transforms of the Fokker-Planck equation, Eq. (5) is expressed in the form:

$$\frac{\partial}{\partial t}p(\xi,v,t) = [\xi\frac{\partial}{\partial v} - \gamma v D_{2-2H}\frac{\partial}{\partial \xi} - \frac{a}{2H\tau^{2H}}t^{2H}\xi v$$
$$-\frac{a}{(2H-1)\tau^{2H-1}}t^{2H-1}v^2]p(\xi,v,t). \quad (7)$$

From now on, we find the solutions of the probability density $P(x,t)$ and $P(v,t)$ in the short-time domain $t << \tau_H$. In order to find two Fourier-transformed solutions for $\xi$ and $v$ by the variable separation from Eq. (7), the two variable-separated equations are given by

$$\frac{\partial}{\partial t}p(\xi,t) = [-\gamma v D_{2-2H}\frac{\partial}{\partial \xi} - \frac{a}{4H\tau_H^{2H}}t^{2H}\xi v - \frac{a}{2(2H-1)\tau_H^{2H-1}}t^{2H-1}v^2 + A]p(\xi,t), \quad (8)$$

$$\frac{\partial}{\partial t}p(v,t) = [\xi\frac{\partial}{\partial v} - \frac{a}{4H\tau_H^{2H}}t^{2H}\xi v - \frac{a}{2(2H-1)\tau_H^{2H-1}}t^{2H-1}v^2 - A]p(v,t). \quad (9)$$

Here, $A$ denotes the separation constant. As we take $\partial p(\xi,t)/\partial t = 0$ in the steady state, we get the steady state probability density $p^{st}(\xi,t)$ for $\xi$ as

$$p^{st}(\xi,t) = \exp[-\frac{a}{2\gamma v D_{2-2H}}[\frac{1}{2H\tau_H^{2H}}t^{2H}v\frac{\xi^2}{2} + \frac{1}{(2H-1)\tau_H^{2H-1}}t^{2H-1}\xi v^2 + A\xi]]. \quad (10)$$

In order to get the solution of the probability density $q^{st}(\xi,t)$ for $\xi$ from $q(\xi,t) \equiv r(\xi,t)q^{st}(\xi,t)$, we calculate the Fourier transform of the probability density after including terms up to order $1/\tau_H^2$ as

E-mail: *kskim@pknu.ac.kr, yjkang66@wku.ac.kr

$$p(\xi,t) = q(\xi,t)\exp[-\frac{a}{2\gamma v D_{2-2H}}[\frac{1}{2H\tau_H^{2H}}t^{2H}v\frac{\xi^2}{2} + \frac{1}{(2H-1)\tau_H^{2H-1}}t^{2H-1}v^2\xi + A\xi]], \quad (11)$$

$$q(\xi,t) = r(\xi,t)\exp[-\frac{a}{2(\gamma v D_{2-2H})^2}[\frac{1}{(2H)^2\tau_H^{2H}}t^{2H}v\frac{\xi^2}{6} + \frac{1}{(2H-1)\tau_H^{2H-1}}t^{2H-1}v^2\frac{\xi^2}{2}]]. \quad (12)$$

Taking the solutions as arbitrary functions of variable $t - \xi/\gamma v D_{2-2H}$, the arbitrary function $r(\xi,t)$ becomes $\Theta[t - \xi/\gamma v D_{2-2H}]$. As a result, we find that

$$p(\xi,t) = r(\xi,t)q^{st}(\xi,t)p^{st}(\xi,t) = \Theta[t - \xi/\gamma v D_{2-2H}]q^{st}(\xi,t)p^{st}(\xi,t). \quad (13)$$

By the similar method of Eqs. (10)-(13) for $v$, we also get the Fourier transform of the probability density for the velocity as

$$p(v,t) = \Theta[t + v/\xi]q^{st}(v,t)p^{st}(v,t). \quad (14)$$

Therefore, by calculating Eq. (13) and Eq. (14), we get the Fourier transform of the joint probability density.

From now on, we find the solutions of the probability density $P(x,t)$ and $P(v,t)$ for the fractional generalized Langevin equation in the three-time domains (a) $t \ll \tau_H$, (b) $t \gg \tau_H$, and (c) $\tau_H = 0$.

(*a*) *short-time domain* $t \ll \tau_H$

We get the probability density $P(x,t)$ and $P(v,t)$ from Eq. (8) and Eq. (9) as

$$p(x,t) = [2\pi \frac{at^{2H+2}}{3(2H-1)\tau_H^{2H-1}}]^{-1/2}\exp[-\frac{3(2H-1)\tau_H^{2H-1}}{2at^{2H+2}}x^2], \quad (15)$$

$$p(v,t) = [2\pi \frac{at^{2H+2}}{(2H-1)\tau_H^{2H-1}}]^{-1/2}\exp[-\frac{(2H-1)\tau_H^{2H-1}}{2at^{2H+2}}v^2]. \quad (16)$$

The mean squared displacement and the mean squared displacement for $P(x,t)$ and $P(v,t)$ are, respectively, given by

$$<x^2(t)> = \frac{a}{3(2H-1)\tau_H^{2H-1}}t^{2H+2}, \quad (17)$$

$$<v^2(t)> = \frac{a}{(2H-1)\tau_H^{2H-1}}t^{2H+2}. \quad (18)$$

(*b*) *long-time domain* $t \gg \tau_H$

By using the inverse Fourier transform, the probability density $P(x,t)$ and $P(v,t)$ are, respectively, presented by

$$p(x,t) = [\pi \frac{at^{2H+2}}{H(2H+1)\tau_H^{2H}}]^{-1/2}\exp[-\frac{H(2H+1)\tau_H^{2H}}{at^{2H+2}}x^2], \quad (19)$$

$$p(v,t) = [\pi \frac{a\gamma t^{2H+1}}{H(2H-1)\tau_H^{2H-1}}]^{-1/2}\exp[-\frac{H(2H-1)\tau_H^{2H-1}}{a\gamma t^{2H+1}}v^2]. \quad (20)$$

The mean squared values $<x^2(t)>$ and $<v^2(t)>$ for the probability density $P(x,t)$ and $P(v,t)$ are, respectively, given by

$$<x^2(t)> = \frac{a}{2H(2H+1)\tau_H^{2H}}t^{2H+2}, \quad (21)$$

$$<v^2(t)> = \frac{a\gamma}{2H(2H-1)\tau_H^{2H-1}}t^{2H+1}. \quad (22)$$

(*c*) *time domain* $\tau = 0$

The probability density $P(x,t)$ and $P(v,t)$ are, respectively, presented by

$$p(x,t) = [2\pi \frac{at^{2H+2}}{3(2H-1)\tau_H^{2H-1}}]^{-1/2}\exp[-\frac{3(2H-1)\tau_H^{2H-1}}{2at^{2H+2}}x^2], \quad (23)$$

$$p(v,t) = [2\pi \frac{at^{2H}}{(2H-1)\tau_H^{2H-1}}]^{-1/2}\exp[-\frac{(2H-1)\tau_H^{2H-1}}{2at^{2H}}v^2]. \quad (24)$$

The mean squared displacement and the mean squared velocity are, respectively, given by

$$<x^2(t)> = \frac{a}{3(2H-1)\tau_H^{2H-1}}t^{2H+2}, \quad <v^2(t)> = \frac{a}{(2H-1)\tau_H^{2H-1}}t^{2H}. \quad (25)$$

In summary, we mainly have derived the Fokker-Planck equation and briefly use the Fourier transform of the joint probability density of the fractional generalized Langevin equation in order to find the joint probability density in the three-time domains (*a*)-(*c*).

As the active particle with a thermal equilibrium noise uses the similar method[18], the mean squared displacement and the mean squared velocity in long-time domain $t \ll \tau_H$ behave typically as $<x^2(t)> \sim t^{2H+2}$ and $<v^2(t)> \sim t^{2H+2}$. Particularly, the mean squared velocity has a sub-diffusive form as $<v^2(t)> \sim t^{2H}$ in the time domain $\tau_H = 0$ or $t \to \infty$, while that exhibits a super-diffusion as $<v^2(t)> \sim t^{2H+1}$ in $t \gg \tau_H$. The active mean squared displacement $<x^2(t)> \sim t^3$ at $H = 1/2$ in the limit of $t \ll \tau_H$ is consistent with that of Ref. [19].

Viscoelastic active diffusion results show some variation from other findings. Future use of the active fraction generalization Langevin equation with thermal noise in our model should improve really the interpretability of results by other models. The non-equilibrium thermodynamic quantities, meticulously calculated from the joint probability density using numerical methods, await validation through comparison with simulation and experimental data.

This paper was supported by Wonkwang University in 2024.